\documentclass[12pt,a4paper]{article}
\usepackage[dvips]{graphicx}
\usepackage{amssymb}
\usepackage{amsmath}
\usepackage{cite}
\usepackage{color}
\newcommand{\bea}{\begin{equation}}
\newcommand{\eea}{\end{equation}}
\newcommand{\be}{\begin{eqnarray}}
\newcommand{\ee}{\end{eqnarray}}
\newcommand{\nn}{\nonumber}
\newcommand{\ev}{\mbox{eV}}
\newcommand{\mev}{\mbox{MeV}}
\newcommand{\gev}{\mbox{GeV}}
\newcommand{\tev}{\mbox{TeV}}
\def\hbar#1{\backslash\hspace{-2mm}#1}

\def\nn{\nonumber}

\def\2tvec#1#2{
\left(
\begin{array}{c}
#1  \\
#2  \\
\end{array}
\right)}

\def\mat2#1#2#3#4{
\left(
\begin{array}{cc}
#1 & #2 \\
#3 & #4 \\
\end{array}
\right) }

\def\Mat3#1#2#3#4#5#6#7#8#9{
\left(
\begin{array}{ccc}
#1 & #2 & #3 \\
#4 & #5 & #6 \\
#7 & #8 & #9 \\
\end{array}
\right) }

\def\3tvec#1#2#3{
\left(
\begin{array}{c}
#1  \\
#2  \\
#3  \\
\end{array}
\right)}

\def\hbar#1{\backslash\hspace{-2mm}#1}

\numberwithin{equation}{section}

\begin{document}

\begin{titlepage}
\begin{flushright}
KIAS-P13022\\
IPPP/13/24\\
DCPT/13/48
\end{flushright}

\begin{center}

\vspace{1cm}
{\large\bf New Interpretation of the Recent Result of  AMS-02 \\ 
\vspace{-1.5mm}and\\ \vspace{-1.5mm} 
 Multi-component Decaying Dark Matters\\ with
non-Abelian Discrete Flavor Symmetry}
\vspace{1cm}

Yuji Kajiyama,$^{a,}$\footnote{kajiyama-yuuji@akita-pref.ed.jp}
Hiroshi Okada,$^{b,}$\footnote{hokada@kias.re.kr}
Takashi Toma$^{c,}$\footnote{takashi.toma@durham.ac.uk}
\vspace{5mm}

{\it%
$^{a}${Akita Highschool, Tegata-Nakadai 1, Akita, 010-0851, Japan}\\
$^{b}${School of Physics, KIAS, Seoul 130-722, Korea}\\
$^{c}${Institute for Particle Physics Phenomenology University of Durham, Durham DH1 3LE, UK}
}

\vspace{8mm}

\abstract{
Recently the AMS-02 experiment has released the data of positron
 fraction with much small statistical error. 
 Because of the small error, it is no longer easy to fit the data with a
 single dark matter for a fixed diffusion model and dark matter profile. 
 In this paper, we propose a new interpretation of the data that 
it originates from decay of two dark matter. 
This interpretation gives a rough threshold of the lighter DM component. 
When DM decays into leptons, 
the positron fraction in the cosmic ray depends on the 
flavor of the final states, and 
 this is fixed by imposing
 non-Abelian discrete symmetry in our model.
By assuming two gauge-singlet fermionic decaying DM particles,
 we show that a model with non-Abelian discrete flavor symmetry, {\it
 e.g.} $T_{13}$, can give a much better fitting to the AMS-02 data
 compared with single dark matter scenario. Few dimension six operators
 of universal leptonic decay of DM particles are allowed in our model since its
 decay operators are constrained by the $T_{13}$ symmetry. 
We also show that the lepton masses and mixings are consistent with current 
experimental data, due to the flavor symmetry.  
}

\end{center}
\end{titlepage}

\setcounter{footnote}{0}

\section{Introduction}

The latest experiment of Planck~\cite{planck} tells us that  about 26.8
$\%$ of energy density of the universe consists of Dark Matter (DM). 
Many experiments are being performed to search DM signatures. 
The recent result of the indirect detection experiment of
AMS-02~\cite{ams-02} is in favor of the previous experiments such as
PAMELA~\cite{Adriani:2008zr,Adriani:2011cu} and Fermi-LAT~\cite{fermi}, which had reported
the excess of positron fraction in the cosmic ray. Moreover, it smoothly
extends the anomaly line of positron fraction with energy up to about
350 GeV with small statistics error compared with the previous
experiments. 
These observations can be, in general, explained by scattering and/or 
decay of the GeV/TeV-scale DM particles.  
In addition leptophilic DM is preferable since PAMELA observed no anti-proton
excess \cite{Adriani:2008zq}. 
Along this line of thought, several papers have been released
\cite{Yin:2013vaa,Yuan:2013eba,Kopp:2013eka,DeSimone:2013fia,Yuan:2013eja,
Ibe:2013nka,Linden:2013mqa,Cholis:2013psa,Jin:2013nta}. 
Due to the smallness of statistics error of AMS-02, it became difficult
to fit to the data as same as the previous experiments like PAMELA~\cite{Cholis:2013psa}. 

In this paper, we show that we can obtain a better fitting to the data
with two component decaying DM. We introduce two
kinds of fermionic DM particles with mass of ${\cal O}(100)$ GeV and
${\cal O}(1)$ TeV into the framework of $T_{13}$ flavor symmetric
model~\cite{t13-origin}. 
In our model, the flavor symmetry $T_{13}$ works at least in two ways:
(i) It constrains interactions between DM and the Standard Model (SM)
particles. DM particles which are gauge singlet fermion $X$ and
$X'$ couple with leptons by dimension six operators $\bar L E \bar L
X^{(')}/\Lambda^2$ due to the $T_{13}$ symmetry, thus these are
leptophilic. DM particles decay into leptons via these operators with
the suppression factor $\Lambda\sim 10^{16}~\gev$, giving  desired
lifetime of DM particles, $\Gamma^{-1} \sim
((\tev)^5/\Lambda^4)^{-1}\sim 10^{26}$
s\cite{a4-decay,a5-prime-decay}. (ii) Flavor of the final states of DM
decay is determined by the $T_{13}$ symmetry. We give a
concrete example of the universal final states
$X/X'\to\nu_ee^+e^-/\nu_\mu\mu^+\mu^-/\nu_\tau\tau^+\tau^-$.  
Due to a specific selection rule by the flavor symmetry mentioned above, 
we show that two-component DM model is preferable for explanation of 
the precise AMS-02 result. 
In addition to that, we find a set of parameters that is consistent with the observed 
lepton masses and their mixings especially somewhat large angle of $\theta_{13}$ 
recently reported by several experiments \cite{t2k, reno, daya-bay, double-chooz, 
chooz, pvc, minos}.

This paper is organized as follows. In Section~\ref{sec:T7}, we briefly mention
the $T_{13}$ symmetric model and construct mass matrices of the lepton
sector in definite choice of $T_{13}$ assignment of the fields. We show
that there exists a consistent set of parameters. 
In section~\ref{sec:dm}, we show that  
desirable dimension six DM decay operators are allowed by $T_{13}$ 
symmetry and that leptonic decay of the two DM particles by those
operators shows good agreement with the cosmic-ray anomaly
experiments. Section~\ref{sec:conc} is devoted to the conclusions.    
%%%%%%%%%%%%%%%%%%%%%%%%%%%%%%%%%%%%%%%%

\section{Lepton masses and mixings with $T_{13}$ flavor symmetry }
\label{sec:T7}

First of all, we briefly review our model based on the non-Abelian
discrete group $T_{13}$, which is isomorphic to $Z_{13}\rtimes
Z_3$~\cite{t13-origin,Fairbairn:1982jx,King:2009ap,review}. 
The $T_{13}$ group is a subgroup of $SU(3)$, and known as the minimal
non-Abelian discrete group having two complex triplets as the
irreducible representations, see ref.~\cite{t13-origin} for details. 

Lepton masses and mixings are derived from the setup shown in
Table~\ref{T13-assign}. Here, $Q$, $U$, $D$, $L$, $E$, $H (H')$ and
$X(X')$ denote left-handed quarks, 
right-handed up-type quarks, right-handed down-type quarks, left-handed leptons, 
right-handed charged leptons, Higgs bosons, and gauge singlet fermions,
respectively\footnote{All the assignment and particle contents are the
same as our previous work~\cite{t13-origin} except the DM sector.}. 
Here one should notice that $X$ and $X'$ is Majorana- and Dirac-type DM,
respectively that directly comes from the charge assignment of $T_{13}$. 
%%%
Due to the  $T_{13}$ flavor symmetry in addition to an appropriate choice of 
the additional $Z_3$ symmetry,  triplet Higgs bosons $H({\bf 3_1})$ and 
$H({\bf \bar 3_2})$ couple only to leptons, while $T_{13}$ singlet Higgs bosons 
$H' ({\bf 1_{0,1,2}})$ couple only to quarks. Hence a linear combination of $H'$ 
is the SM-like Higgs boson and is created at LHC by gluon fusion. 
Therefore, the mass matrices of quark sector are not constrained, while
those of lepton sector are determined by the $T_{13}$ symmetry. 
For the neutrino sector, since the Yukawa couplings $LHX$ and $LHX'$ are 
forbidden by the $T_{13}$ symmetry, 
the left-handed Majorana neutrino mass terms are derived from dimension five operators $LHLH$. 
Here notice that $X$ and $X'$ have dimension six operators $\bar LE \bar
LX$, $\bar LELX'$, and mass terms $m_X XX$, $m_{X'} \bar X' X' $. 
%%%
For the matter content and the $T_{13}$ assignment given in
Table~\ref{T13-assign}, the charged-lepton and neutrino masses are generated 
from the $T_{13}$ invariant operators
\be
{\cal L}_{Y}&=&\sqrt{2}a_e \bar E L H^c ({\bf \bar 3_2})+
\sqrt{2}b_e \bar E L H^c ({\bf 3_1})\nn\\
&&+\frac{a_{\nu}}{\Lambda}LH({\bf \bar 3_2})LH({\bf \bar 3_2})
+\frac{b_{\nu}}{\Lambda}(LH({\bf \bar 3_2}))_{\bf \bar 3_2}
(LH({\bf 3_1}))_{\bf 3_2}\nn\\
&&+\frac{c_{\nu}}{\Lambda}(LH({\bf \bar 3_2}))_{\bf 3_1}
(LH({\bf 3_1}))_{\bf \bar 3_1}+\mathrm{h.c.},
\label{lag}
\ee
where $H^c=\epsilon H^*$, and 
$LH({\bf \bar 3_2})LH(\bf 3_1)$ is $T_{13}$ invariant in two different products, 
corresponding to $b_{\nu}$ and $c_{\nu}$.  
The fundamental scale
$\Lambda=10^{11}~\gev$ is needed for the certain neutrino mass scale
($\Lambda/\sqrt{\lambda}\sim 10^{16}$ GeV is 
required to obtain the desired lifetime of DM, where $\lambda$ is
coupling constant of DM decay operators as we will discuss later). 
After the electroweak symmetry breaking, the Lagrangian Eq.~({\ref{lag}})
gives rise to mass matrices of charged leptons $M_e$ and neutrinos $M_{\nu}$ as
\be
M_e&=&\left( \begin{array}{ccc} 
0& b_e v_1 & a_e \bar v_2\\
a_e \bar v_3 &0& b_e v_2 \\
b_e v_3 &a_e \bar v_1&0 \\
\end{array}\right), ~
\label{me}\\
M_{\nu}&=&
%{\tiny
\frac{1}{\Lambda}\left( \begin{array}{ccc} 
c_{\nu}\bar v_3 v_2&a_{\nu} \bar v_1^2+b_{\nu}\bar v_3 v_1&a_{\nu}\bar v_3^2+b_{\nu}\bar v_2 v_3\\
a_{\nu} \bar v_1^2+b_{\nu}\bar v_3 v_1& c_{\nu}\bar v_1 v_3&a_{\nu}\bar v_2^2+b_{\nu}\bar v_1 v_2 \\
a_{\nu}\bar v_3^2+b_{\nu}\bar v_2 v_3&a_{\nu}\bar v_2^2+b_{\nu}\bar v_1 v_2&c_{\nu} \bar v_2 v_1\\
\end{array}\right)%} 
, 
\nn\\
\label{mnu}
\ee
where the vacuum expectation values (VEVs) of the Higgs bosons are defined as 
$\langle H({\bf 3_1})^i \rangle={v_i}/{\sqrt{2}}$, 
$\langle H({\bf \bar3_2})^i \rangle={\bar v_i}/{\sqrt{2}}$,
$\langle H'({\bf 1_{0,1,2}})\rangle={v'_i}/{\sqrt{2}}$,
$\sum_{i=1}^3 \left( v_i^2+\bar v_i^2+{v'}_{i-1}^2\right)=(246~\gev)^2$. 

\begin{table}[t]
\centering
%\begin{tiny}
\begin{tabular}{c|ccccccccc} \hline\hline
& $Q$ & $U$ & $D$ & $L$ & $E$ & $H$ &$H'$& $X$ & $X'$ \\ \hline
$SU(2)_L \times U(1)_Y$ & 
{\bf 2}$_{1/6}$  & {\bf 1}$_{2/3}$ & {\bf 1}$_{-1/3}$ &
~{\bf 2}$_{-1/2}$~ & ~{\bf 1}$_{-1}$~  & {\bf 2}$_{1/2}$ &  ~{\bf 2}$_{1/2}$ &
~{\bf 1}$_0$~&~{\bf 1}$_0$~ \\
$T_{13}$ & 
${\bf 1_{0,1,2}}$ & ${\bf 1_{0,1,2}}$ & ${\bf1_{0,1,2}}$ &
${\bf 3_1}$ & ${\bf 3_2}$ &
${\bf 3_1},{\bf \bar 3_2}$ &${\bf1_{0,1,2}}$ &
${\bf 1_0}$ & ${\bf 1_1}$ \\ 
$Z_3$ & 
$1$ & $\omega$ & $\omega^2$ &
$1$ & $1$ & $1$ &$\omega$ & $1$ & $1$\\ 
\hline
\end{tabular}
\caption{\small The $T_{13}$ and $Z_3$ charge assignment of the SM fields and the Majorana DM $X$ and the Dirac DM $X'$, where $\omega=e^{2i \pi/3}$. }
\label{T13-assign}
%\end{tiny}
\end{table}
%%%

 Now we give a numerical example. By the following choice of parameters, 
\be
v_1&=&0.4269~\gev,~v_2=16.11~\gev,~v_3=7.862~\gev,\nn\\
\bar v_1&=&1~\gev,~\bar v_2=16.82~\gev,~\bar v_3=0.004836~\gev,\nn\\ 
a_e&=&0.1057,~b_e=0,~a_{\nu}=-8.220\times 10^{-3},\nn\\
b_{\nu}&=&8.439 \times 10^{-3},~c_{\nu}=3.632\times 10^{-1},
\label{vabc}
\ee
the mass matrices Eq.~(\ref{me}) and (\ref{mnu}) give rise to mass eigenvalues and 
related observables as
\be
&&m_e = 0.511~\mev,~m_{\mu}=105.7~\mev,~m_{\tau}=1777~\mev,\nonumber\\
&&m_{\nu 1}=6.324 \times 10^{-3}~\ev,~m_{\nu 2}=1.078 \times
10^{-2}~\ev,~m_{\nu 3}=5.046 \times 10^{-2}~\ev, \nn\\ 
&&\Delta m_{21}^2=m_{\nu 2}^2-m_{\nu 1}^2=7.62  \times 10^{-5}~\ev^2,\nn\\
&&\Delta m_{32}^2=m_{\nu 3}^2-m_{\nu 2}^2=2.43 \times 10^{-3}~\ev^2,~ \\
&&\langle m\rangle_{ee}=2.83 \times 10^{-4}~\ev,~\sum_i m_{\nu i}=5.49 \times 10^{-2}~\ev,\nn
\ee
and the mixing matrices are given by
\be
U_{eL}&=&\left( \begin{array}{ccc}
1& 0&0\\
0&1&0\\
0&0&1\\
\end{array}\right),~
U_{eR}=\left( \begin{array}{ccc}
0& 0&1\\
1&0&0\\
0&1&0\\
\end{array}\right),~\label{uelr}\\ \nn
U_{MNS}&=&U_{eL}^{\dag}U_{\nu}=\left( \begin{array}{rrr}
%0.818831 & 0.552308&-0.156434\\
%-0.303704&0.648074&0.698401\\
%-0.487114&0.524362&-0.698401\\
0.819 & 0.552&-0.156\\
-0.304&0.648&0.698\\
-0.487&0.524&-0.698\\
\end{array}\right),\\
\theta_{12}&=&34^\circ,~\theta_{23}=-45^\circ,~\theta_{13}=-9^\circ,
\ee
which are all consistent with the present experimental data
\cite{pdg,Tortola:2012te}. In particular in the case of $U_{eL}=1$, 
the mass matrices Eq.~(\ref{me}) and  (\ref{mnu}) require 
normal hierarchy $m_{\nu 1}<m_{\nu 2}<m_{\nu 3}$ of the neutrino masses
and $(U_{MNS})_{e3}\neq 0$.
A comprehensive analyses of the $T_{13}$ symmetric models have been done
by several authors  \cite{Parattu:2010cy, Ding:2011qt, Hartmann:2011pq,
Hartmann:2011dn}. Although one can sweep whole range of 
parameters, we adopt those of Eq.~(\ref{vabc}) giving universal final states due to 
the mixing matrices Eq.~(\ref{uelr}) since such analysis is out of scope 
of the present paper.

As for the Higgs sector, since the present model contains nine Higgs doublets, 
it causes flavor changing neutral current processes such as $\bar K^0-K^0$ mixings. 
Therefore, extra Higgs bosons must be enough heavy. 
Moreover, additional massless bosons appear because the $T_{13}$ symmetric Higgs 
potential has accidental $U(1)$ symmetry. Therefore one can introduce soft $T_{13}$ 
breaking terms such as $H'^{\dag}_{\bf 1_0}H'_{\bf 1_1}+
H'^{\dag}_{\bf 1_0}H'_{\bf 1_2}+H'^{\dag}_{\bf 1_2}H'_{\bf 1_0}$ and 
$H^{\dag}_{\bf 1_0}\sum_i H(\bar 3_2)^i$ in order to avoid those problems.

%%%%%%%%%%%%%%%%%%%%%%%%%%%%%%%%%%%%%%%
\section{Decaying dark matter in the $T_{13}$ model}
\label{sec:dm}
%%%%%%%%%%%%%%%%%%%%%%%%%
\begin{table}[h]
\centering
\begin{tabular}{ccl} \hline\hline
Dimensions && \multicolumn{1}{c}{DM decay operators} \\ \hline
4 && $\bar{L} H^c X^{(')}$ \\ 
5 && ~~~$-$ \\
6 && 
$\bar{L}E\bar{L}X^{(')}$,
~~$H^\dagger\!H\bar{L}H^cX^{(')}$,
~~$(H^c)^tD_\mu H^c\bar{E}\gamma^\mu X^{(')}$, \\
&&
$\bar{Q}D\bar{L}X^{(')}$,
~~$\bar{U}Q\bar{L}X^{(')}$,
~~$\bar{L}D\bar{Q}X^{(')}$,
~~$\bar{U}\gamma_\mu D\bar{E}\gamma^\mu X^{(')}$, \\
&& 
$D^\mu H^c D_\mu \bar{L} X^{(')}$,
~~$D^\mu D_\mu H^c\bar{L}X^{(')}$,  \\
&& 
$B_{\mu\nu}\bar{L}\sigma^{\mu\nu}H^cX^{(')}$,
~~$W_{\mu\nu}^a\bar{L}\sigma^{\mu\nu}\tau^aH^cX^{(')}$  \\ \hline
\end{tabular}
\medskip
\caption{\small The higher dimensional operators which cause decay of
 $X$ and $X'$ up to dimension six~\cite{delAguila:2008ir}. 
$B_{\mu\nu}$, $W_{\mu\nu}^a$, and $D_\mu$ are the field strength
tensors of hypercharge gauge boson, weak gauge boson, and the
electroweak covariant derivative.\bigskip}
\label{op}
\end{table}

%%%
It is well known that the cosmic-ray anomalies measured by
PAMELA~\cite{Adriani:2008zr} and Fermi-LAT~\cite{fermi}
can be explained by DM decay with lifetime of $\Gamma^{-1} \sim 10^{26}$
s. If the DM ($X$ and $X'$ in our case)
decays into leptons by dimension six operators $\bar L E \bar L
X^{(')}/\Lambda^2$ with $\Lambda\sim10^{16}~\mathrm{GeV}$, such long
lifetime can be achieved.  
In general, however, there exist several gauge invariant decay operators
of lower dimensions; dimension four operators inducing too
rapid DM decay, and dimension six operators including quarks, Higgs and gauge
bosons in the final states, which must be forbidden in a successful
model.
%%%
By the field assignment of Table~\ref{T13-assign}, most of all the decay
operators listed in Table~\ref{op}~\cite{delAguila:2008ir} 
are forbidden due to the $T_{13}$ symmetry except for
$\bar{L}E\bar{L}X^{(')}
%%%%%%%
$\footnote{
Notice that
$H^\dagger\!H\bar{L}H^cX^{(')}$ and 
$H^\dagger HXX^{'}$ cannot be forbidden by any symmetries that hold unitarity.
Moreover, these interactions 
induce decay of one DM to the other DM. Here we assume these couplings
of these surviving  terms to be enough tiny.
The most stringent constraint process is $X'\to X,h$ that  comes from
$H^\dagger HXX^{'}$, where $h$ is the standard model Higgs whose mass is
126 GeV~\cite{Higgs}. We find that its coupling should be less than
${\cal O}(10^{-18})$ in order to conservatively satisfy the no excess
constraint of the antiproton with the lifetime of DM to be longer than
${\cal O}(10^{28})$ s. 
%However, we expect that these terms can be forbidden by modifying the other flavor symmetries
%with higher order elements such as $\Sigma$(81) of $\Sigma(3N^3)$~\cite{ishi-koba}, $A_5$~\cite{a5}, the double covering of
%$A_5$~\cite{a5-prime-decay}, and $T_{19}$ of $T_N$.
}
%%%%%
. 
Therefore, one do not have to be worried about production of anti-proton
and secondary positron by scattering with nucleon and interstellar
medium. 
With the notation $L_i=(\nu_{i},\ell_{i})=(U_{eL})_{i
\alpha}(\nu_{\alpha},\ell_{\alpha})$ 
and $E_i=(U_{eR})_{i \beta}E_{\beta}$
$(i=1,2,3,~\alpha,\beta=e,\mu,\tau)$, 
the four-Fermi decay interaction is explicitly written as 
\begin{eqnarray}
{\cal L}_{\rm decay}
&=&
\frac{\lambda_X}{\Lambda^2}\,\sum_{i=1}^3(\bar{L}_iE_i)\bar{L}_iX
+\frac{\lambda_{X'}}{\Lambda^2}\sum_{i=1}^{3}\left(\omega^{2(i-1)}\right)
\left(\bar{L}_iE_i\right)\bar{L}_iX'+\text{h.c.} \nn\\
&=&\frac{\lambda_X}{\Lambda^2}\sum_{i=1}^3 
\sum_{\alpha, \beta,\gamma}
\left( U_{eL}\right)_{i \alpha}^* \left( U_{eR}\right)_{i\beta}\left(
 U_{eL}\right)_{i \gamma}^* 
\label{lag2} \nn\\
&&\times\left[ \left( \bar \nu_{\alpha}P_R E_{\beta}\right)\left( \bar
		\ell_{\gamma}P_R X \right)
-\left( \bar\ell_{\gamma}P_R E_{\beta}\right)\left( \bar \nu_{\alpha}P_R
  X \right)\right]
  +\left(X\to X'\right)+\text{h.c.},
\end{eqnarray}
where the factor $\left( \omega^{2(i-1)}\right)$ is only for the case of 
$X'$ decay
because of the multiplication rule of the $T_{13}$ flavor symmetry.
As seen from Eq.~(\ref{lag2}), decay mode of the DM particles $X$ and $X'$
depends on the mixing matrices $U_{eL}$ and $U_{eR}$, which are given in
Eq.~(\ref{uelr}). 

Next, we consider the decay width of the decaying DM 
through the $T_{13}$ invariant interaction Eq.~(\ref{lag2}). Due to the
particular generation structure, the DM particles $X$ and $X'$ decay into
several tri-leptons final state with the mixing-dependent rate.
The decay width of DM $X$ per each flavor is defined as 
$\Gamma_{\alpha\beta\gamma}\equiv
 \Gamma(X\to\nu_\alpha\ell_\beta^+\ell_\gamma^-)
+\Gamma(X\to\overline{\nu_\alpha}\ell_\beta^+\ell_\gamma^-)$,
and the decay width $\Gamma_{\alpha\beta\gamma}$ is calculated as
\begin{equation}
\Gamma_{\alpha\beta\gamma}
=\frac{\left|\lambda_X\right|^2m_X^5}{32\left(4\pi\right)^3\Lambda^4}\left(U_{\alpha\beta\gamma}
+U_{\alpha\gamma\beta}\right),
\label{decaywidth}
\end{equation}
where
\begin{equation}
U_{\alpha \beta \gamma}=
\left| \sum_{i=1}^3\left( U_{eL}\right)_{i
 \alpha}^* \left( U_{eR}\right)_{i\beta}\left( U_{eL}\right)_{i
 \gamma}^*\right|^2. 
\end{equation}
The decay width of $X'$ named $\Gamma_{\alpha\beta\gamma}'$ is obtained
by replacing $X\to X'$. % and multiplying the factor $\left(\omega^{2(i-1)}\right)$. 
The differential decay width is written as
\begin{equation}
\frac{d\Gamma_{\alpha\beta\gamma}}{dx}=
\frac{\left|\lambda_X\right|^2m_X^5}{48\left(4\pi\right)^3\Lambda^4}
x^2\Bigl((6-2x)U_{\alpha\beta\gamma}+
\left(15-14x\right)U_{\alpha\gamma\beta}\Bigr),
\label{eq:ddecaywidth}
\end{equation}
where $x=2E_{\ell_{\beta}^+}/m_X$. This is required to calculate the energy distribution
function of injected $e^\pm$ from DM decay, $dN_{e^\pm}/dE$. 
Here we have neglected the masses of charged leptons in the final states. 
In both $X$ and $X'$ DM cases, the flavor dependent factor $U_{\alpha
\beta \gamma}$ gives a factor three when one takes the sum of flavor
indices $\alpha,\beta$ and $\gamma$. That is not by a particular choice
of parameters Eq.~(\ref{vabc}), but by the $T_{13}$ symmetry. 
Therefore, the branching fraction of each decay mode is given by
%\begin{eqnarray}
%&&\!\!\!\!
$\mathrm{Br}(X \to
\nu_{\alpha}\ell^+_{\beta}\ell^-_{\gamma},~\overline{\nu_{\alpha}}\ell^+_{\beta}\ell^-_{\gamma}) 
=\left( U_{\alpha \beta \gamma}+U_{\alpha\gamma \beta}\right)/6.
$
%\end{eqnarray}
The DM mass $m_{X}$ and the total decay width 
$\Gamma_X=\sum_{\alpha,\beta,\gamma}\Gamma_{\alpha\beta\gamma}$ 
are chosen to be free parameters in the following analysis since it can
be always tuned with the coupling $\lambda_X$ and the cut-off scale $\Lambda$. 
%In the case of $X'$, one can obtain a similar formulae by replacing $X\to X'$. 

Given the differential decay width and the branching ratios, the primary
source term of the positron and electron coming from DM decay at the 
position $r$ of the halo associated with our galaxy is expressed as
\begin{eqnarray}
 Q(E,r) \!\!\!&=&\!\!\! n_{X}(r) \,\Gamma_{X}
\sum_f {\rm Br}(X \to f) \bigg(\frac{dN_{e^\pm}}{dE}\bigg)_f%\nonumber\\&&\!\!\!
+\left(X\to X'\right),
\label{source}
\end{eqnarray}
where $\left(dN_{e^\pm}/dE\right)_f$ is the energy distribution of
$e^\pm$ coming from the DM decay with the final 
state $f$, and $E$ is the energy of injected $e^{\pm}$. 
We use the PYTHIA 8~\cite{Sjostrand:2006za} 
to evaluate the energy distribution function.
%%%%%%%
Although it is often assumed that the relic density of the DM is
thermally determined, non-thermal production of the DM dark matter is
also possible~\cite{non-thermal}. We thus do not specify the origin of
the relic DM in the following analysis, and assume that the number
densities of $X$ and $X'$ are the same for simplest case. 
%%%%%%%
The non-relativistic DM number density $n_{X}(r)$ is rewritten by 
$n_{X}(r)=\rho_{X}(r)/m_{X}$ with the DM profile $\rho(r)$. 
In this work, we adopt the
Navarro-Frank-White (NFW) profile~\cite{Navarro:1996gj},
\begin{eqnarray}
  \rho_{\rm NFW}(r) \,=\,
  \rho_\odot\frac{r_\odot(r_\odot+r_c)^2}{r(r+r_c)^2},
\end{eqnarray}
where $\rho_\odot\simeq0.40$~GeV/cm$^3$ is the local DM density
at the solar system, $r$ is the distance from the galactic center
whose special values $r_\odot\simeq8.5$~kpc and $r_c\simeq 20$~kpc
are the distance to the solar system and the core radius of the
profile, respectively.
%%%%%%%%%%%%%%%
The diffusion equation must be solved to evaluate the $e^\pm$ flux
observed at the Earth, and it depends on diffusion model. 
The observable $e^\pm$ flux at the solar system $d\Phi_{e^\pm}/dE$ which is produced by DM
decay is given by 
\begin{equation}
\frac{d\Phi_{e^\pm}}{dE}=\sum_{X,X'}\frac{v_{e^\pm}}{4\pi
 b(E)}\frac{\rho_\odot}{m_X}\Gamma_X
\sum_{f}\mathrm{Br}(X\to f)
\int_{E}^{m_X}\left(\frac{dN_{e^\pm}}{dE'}\right)_fI_\odot\left(E,E'\right)dE',
\end{equation}
where $b(E)$ is a space-independent energy loss coefficient written as
$b(E)=E^2/(\tau_{\odot}\cdot1\mathrm{GeV})$ with
$\tau_\odot=5.7\times10^{15}~\mathrm{s}$, and $I_\odot(E,E')$ is
the reduced halo function at the solar system which is expressed by
Fourier-Bessel expansion~\cite{Cirelli:2010xx}. A fitting function for 
the reduced halo function $I(\lambda_D)$ is given in ref.~\cite{Cirelli:2010xx} as a
function of a single parameter $\lambda_D$ which is called diffusion
length. The diffusion length $\lambda_D$ is given by 
\begin{equation}
\lambda_D^2=\frac{4K_0\tau_\odot}{1-\delta}
\left[E^{\delta-1}-{E'}^{\delta-1}\right],
\end{equation}
where we use the following diffusion parameters: $\delta=0.70$,
$K_0=0.0112~\mathrm{kpc^2/Myr}$ which is called MED. In addition, the diffusion zone is
considered as a cylinder that sandwiches the galactic plane with
height of $2L$ and radius $R$ where $L=4~\mathrm{kpc}$ and 
$R=20~\mathrm{kpc}$.

As seen from Eq. (\ref{uelr}) and (\ref{decaywidth}), the DM decays into 
$e^{\pm}$ as well as $\mu^{\pm}$ and $\tau^{\pm}$ in the equal rate. 
As a result, pure leptonic decays give dominant contributions, and 
it is consistent with no anti-proton excess of the PAMELA
results~\cite{Adriani:2008zq}. 
We may take into account the gamma-ray constraint since a lot of
gamma-ray is produced by the hadronization of $\tau^\pm$. As we
see below, the obtained lifetimes of DM particles $X$ and $X'$ are roughly
$\tau_X,\tau_{X'}\gtrsim5\times10^{26}~\mathrm{s}$. 
Thus we do not need to consider the gamma-ray constraint seriously as
long as comparing with ref.~\cite{Ackermann:2012qk}.

%%%%%%%%%%%%%%%%%%%%%%%%%%
\underbar{Result for AMS-02}
\begin{figure}[t]
\begin{center}
\includegraphics[scale=0.85]{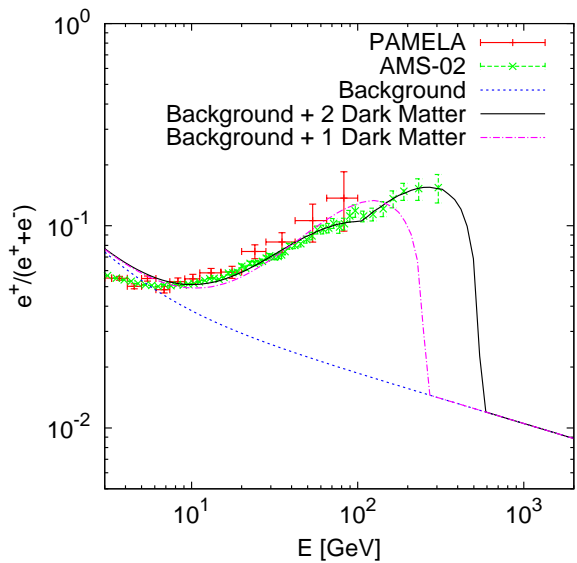}
\caption
{\small The positron fraction \cite{ams-02} and \cite{Adriani:2008zr}
predicted in the leptonically-decaying two-component DM scenario with 
$T_{13}$ symmetry (solid) and single-component scenario (dashed). For
 two-component DM, we have fixed to the best fit point: $m_X=$208 GeV,
 $m_{X'}=$1112 GeV, $\Gamma_X^{-1}=1.9\times10^{27}~\mathrm{s}$ and
 $\Gamma_{X'}=4.7\times10^{26}~\mathrm{s}$.} 
\label{fig:results}
\end{center}
\end{figure}

We use 31 data points of AMS-02 which are higher than 20 GeV for
chi-square analysis. 
The only statistics error is taken into account as the experimental
errors here~\cite{ams-02}. 
The positron fraction for the scenario of the
leptonically decaying DM with $T_{13}$ symmetry is
depicted in Figure~\ref{fig:results} with the experimental data of AMS-02 and PAMELA.
The flux coming from only one-component DM is also shown in the figure
for comparison. 
The obtained best fit point for one-component DM is $m_X=521~\mathrm{GeV}$,
$\Gamma_X^{-1}=5.1\times10^{26}~\mathrm{s}$ with
$\chi_{\mathrm{min}}^2=172.2$ (29 d.o.f.). 
For the single DM, the positron fraction in the high energy cannot be fit well as one can
see from the figure. This is because the experimental data in low energy
region $E\sim20~\mathrm{GeV}$
has much higher precision, and the energy spectrum $dN_{e^\pm}/dE$ is fixed
by the imposed flavor symmetry, thus the predicted flux in the higher
energy region is almost determined by the flavor symmetry. 
One should note that fitting with one-component DM would be better for
different diffusion models or different DM halo profiles since the
evaluated $e^\pm$ flux has a large dependence of them. 

On the other hand, the fitting parameters for two-component DM are 
\begin{eqnarray}
&m_{X}=208~\mathrm{GeV},\quad
&\Gamma_X^{-1}=1.9\times10^{27}~\mathrm{s},\\
&m_{X'}=1112~\mathrm{GeV},\quad
&\Gamma_{X'}^{-1}=4.7\times10^{26}~\mathrm{s},
\end{eqnarray}
with $\chi_\mathrm{min}^2=22.62$ (27 d.o.f) at the best fit point. 
Therefore the much better fitting is obtained with two-component case. 
This is the result of multi-component DM and the fixed flavor of final 
states by $T_{13}$ symmetry. 
That is not by a particular choice of parameters Eq.~(\ref{vabc}),
but by the $T_{13}$ symmetry as mentioned below Eq.~(\ref{eq:ddecaywidth}). 
A sharper drop-off is expected if we have larger branching ratio for
directly produced positron. 
%%%%%%

% \cite{Adriani:2008zr} and Fermi-LAT \cite{Abdo:2009zk,collaboration:2010ij}
%%%%%%%%%%%%%%%%%%%%%%%%%%
\section{Conclusions}
\label{sec:conc}

We revisited a decaying DM model with  a non-Abelian discrete 
symmetry $T_{13}$, and extended it to the two-component DM scenario by
adding an extra DM $X'$. 
We have shown that our model is consistent with all the observed masses
and mixings in the lepton sector. 
 Due to also the specific selection rule of $T_{13}$, we have found that
 both of DM particles have the universal decay coming from dimension six
 operators that gives a promising model in current indirect detection
 searches of DM. 

Fitting to the positron fraction with a single DM under the assumption
of MED diffusion model and NFW DM profile can no longer give a
good interpretation of the positron excess by DM decay because of the precise measurement of
AMS-02. However taking into account two-component DM as our model gives
much better fitting to AMS-02 observation. 
The obtained parameters are $m_X=$208 GeV with
$\Gamma_{X}^{-1}=1.8\times10^{27}$ s and
$m_{X'}=$1112 GeV  with
$\Gamma_{X'}^{-1}=4.7\times10^{26}$~s, assuming that
$X$ and $X'$ have equal number density.

%%%%%%%%%%%%%%%%%%%%%%%%%%%%%%%%%%%
%\newpage
\vspace{0.5cm}
\hspace{0.2cm} {\bf Acknowledgments}
\vspace{0.5cm}

We would like to thank Prof. Shigeki Matsumoto for a crucial advice. 
T.T. acknowledges support from the European ITN project (FP7-PEOPLE-2011-ITN,
PITN-GA-2011-289442-INVISIBLES). The numerical calculations were carried out on SR16000 at YITP
in Kyoto University.

%%%%%%%%%%%%%%%%%%%%%%%%%%%%%%%%%%%

\end{document}